\documentclass[aps,prx,reprint,groupedaddress,floatfix]{revtex4-2}

\usepackage{graphicx,amsmath}

\usepackage{hyperref}\hypersetup{colorlinks=true, citecolor=blue, urlcolor=blue, linkcolor=blue}

\newcommand{\kB}{k_\mathrm{B}}

\begin{document}


\title{Minimal microscopic model for liquid polyamorphism and water-like anomalies}


\author{Frédéric Caupin}
\email[]{frederic.caupin@univ-lyon1.fr}
\affiliation{Institut Lumi\`ere Mati\`ere, Universit\'e de Lyon, Universit\'e Claude Bernard Lyon 1, CNRS, F-69622, Villeurbanne, France}
\author{Mikhaïl A. Anisimov}
\affiliation{Department of Chemical and Biomolecular Engineering and Institute for Physical Science and Technology,\\
University of Maryland, College Park, USA}

\date{September 23, 2021}

\begin{abstract}
Liquid polyamorphism is the intriguing possibility for a single component substance to exist in multiple liquid phases. We propose a minimal model for this phenomenon. Starting with a binary lattice model with critical azeotropy and liquid-liquid demixing, we allow interconversion of the two species, turning the system into a single-component fluid with two states differing in energy and entropy. Unveiling the phase diagram of the non-interconverting binary mixture gives unprecedented insight on the phase behaviors accessible to the interconverting fluid, such as a liquid-liquid transition with a critical point, or a singularity-free scenario, exhibiting thermodynamic anomalies without polyamorphism. The model provides a unified theoretical framework to describe supercooled water and a variety of polyamorphic liquids with water-like anomalies.
\end{abstract}


\maketitle

\section{Introduction}

The Ising model and the lattice-gas model are landmarks in the history of science. They have provided an explanation of phase transitions based on statistical physics and paved the way to the understanding of universality in critical phenomena, one of the greatest achievements in twentieth century physics. The lattice-gas model, although minimal, with sites on a lattice either occupied or empty, captures the essential physics of all fluids near their liquid-vapor critical point. Here we propose a minimal, two-state model for single-component fluids with water-like anomalies.

Water is an everyday liquid, but, for the scientist, it is a puzzling material, which concentrates the largest number of anomalies compared to the ``ordinary'' liquid~\cite{gallo_water_2016}. One intriguing theoretical explanation of these anomalies is ``liquid polyamorphism'' (LP)~\cite{stanley_liquid_2013}, which posits that water may exist under two distinct liquid phases at low temperature. Observing this liquid-liquid transition (LLT) is challenging, because at the required conditions ice is the stable phase, and the liquid phase has a very short lifetime. Nevertheless, a recent study has reported observation of the two liquid phases of water~\cite{kim_experimental_2020}. In addition to the quantum case of superfluidity in helium isotopes, LP has also been reported in experiments on phosphorus~\cite{katayama_macroscopic_2004}, hydrogen~\cite{knudson_direct_2015}, and recently sulfur~\cite{henry_liquid-liquid_2020}. Notwithstanding, LP, in contrast to well-known crystal polymorphism, is still viewed as an exotic and controversial phenomenon. Some atomistic models with a soft repulsion potential demonstrate the possibility of LLT in a pure substance~\cite{xu_relation_2005,vilaseca_isotropic_2011,pinheiro_critical_2017}. A generic, but more phenomenological approach attributes this phenomenon to equilibrium interconversion of two alternative molecular or supramolecular structures~\cite{anisimov_thermodynamics_2018}. Conceptually, this approach resonates with the idea of two competing local structures in cold and supercooled water~\cite{tanaka_simple_2000}.

\begin{figure*}[ttt]
\centerline{\includegraphics[width=0.85\textwidth]{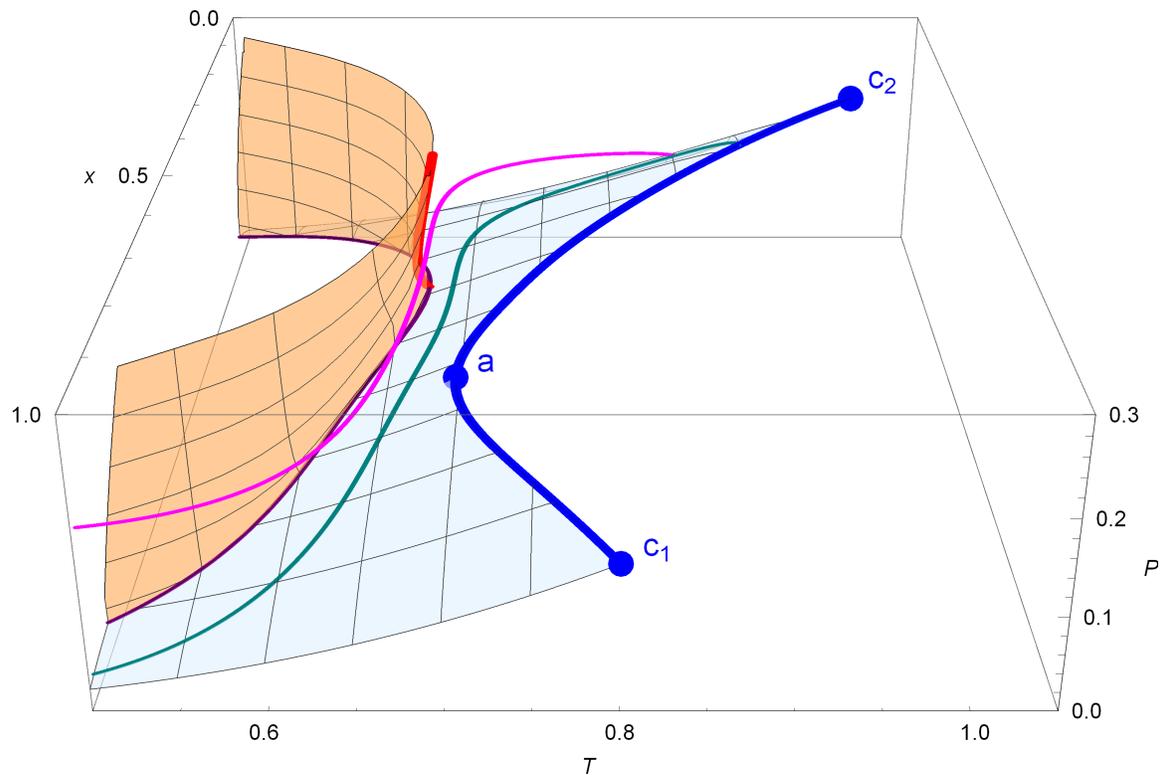}}
\caption{$x-T-P$ phase diagram of a binary mixture with $\omega_1=1.6$, $\omega_2=2$, and $\omega_{12}=1.04$. The light blue surface shows the liquid in equilibrium with its vapor, terminating at the LVcl (dark blue) which reaches its minimum at the critical azeotrope \textbf{a}. The liquid-vapor critical points of the pure components are designated as \textbf{c}$_1$ and \textbf{c}$_2$. The orange surface shows the liquid-liquid equilibrium, terminating at the LLcl (red). The two surfaces intersect along a LLVtl (purple) where two liquids and one vapor coexist. The metastable parts of the liquid-liquid equilibrium surface and critical line are omitted for clarity. When the two species interconvert, the fraction $x$ becomes a function of temperature and pressure, shown as curves for the liquid-vapor (cyan) and the $P=0.18$ isobar (magenta) for $e=3$ and $s=4$.
\label{fig:binary3D}}
\end{figure*}

An equation of state (EoS) based on interconversion of alternative states incorporating a LLT was successfully used to describe the phase behavior and thermodynamic anomalies in two simulated atomistic models of water, ST2~\cite{holten_two-state_2014} and TIP4P/2005~\cite{singh_two-state_2016,biddle_two-structure_2017} or its charge-scaled versions~\cite{horstmann_relations_2021}. A similar EoS was also used to correlate the experimental thermodynamic properties of supercooled water~\cite{caupin_thermodynamics_2019,duska_water_2020} and hydrogen~\cite{cheng_evidence_2020}. Another two-state EoS without a LLT was able to reproduce simulation results for a monatomic water model which does not show LP~\cite{holten_nature_2013}. However, two-state phenomenology still lacks a clear connection with the microscopic nature of underlying inter-molecular interactions. In this work, we mend this gap with a minimalistic lattice model. Previous lattice models~\cite{sastry_singularity-free_1996,hruby_two-structure_2004,ciach_simple_2008,stokely_effect_2010,cerdeirina_waters_2019} were able to produce water-like anomalies, but they were focused on the one-component fluid. Simulations can explore families of model interaction potentials with water-like anomalies by varying a parameter~\cite{gibson_metastable_2006,smallenburg_tuning_2015,dhabal_comparison_2016,russo_water-like_2018,horstmann_relations_2021}, but they are intrinsically confined to the one-component fluid. Here, we go one step back to first display the case when interconversion is absent, making the system a binary fluid. This gives a phase diagram in three dimensions: temperature $T$, pressure $P$, and fraction $x$ of one of the components. Then, we turn interconversion on, which makes $x$ a function of $T$ and $P$, dictated by interconversion equilibrium conditions. The system becomes a one-component fluid, whose phase diagram can be thought as a 2D-manifold immersed in the underlying 3D binary phase diagram. This gives unprecedented insight on the way liquid-liquid polyamorphism may emerge, and provides a general theoretical framework for understanding the variety of possible cases, e.g. with or without LLT, and with opposite signs of the LLT $\mathrm{d}P/\mathrm{d}T$ slopes, such as water and sulfur.

\section{Binary mixture without interconversion of species}
\begin{figure*}[ttt]
\centerline{\includegraphics[width=0.87\textwidth]{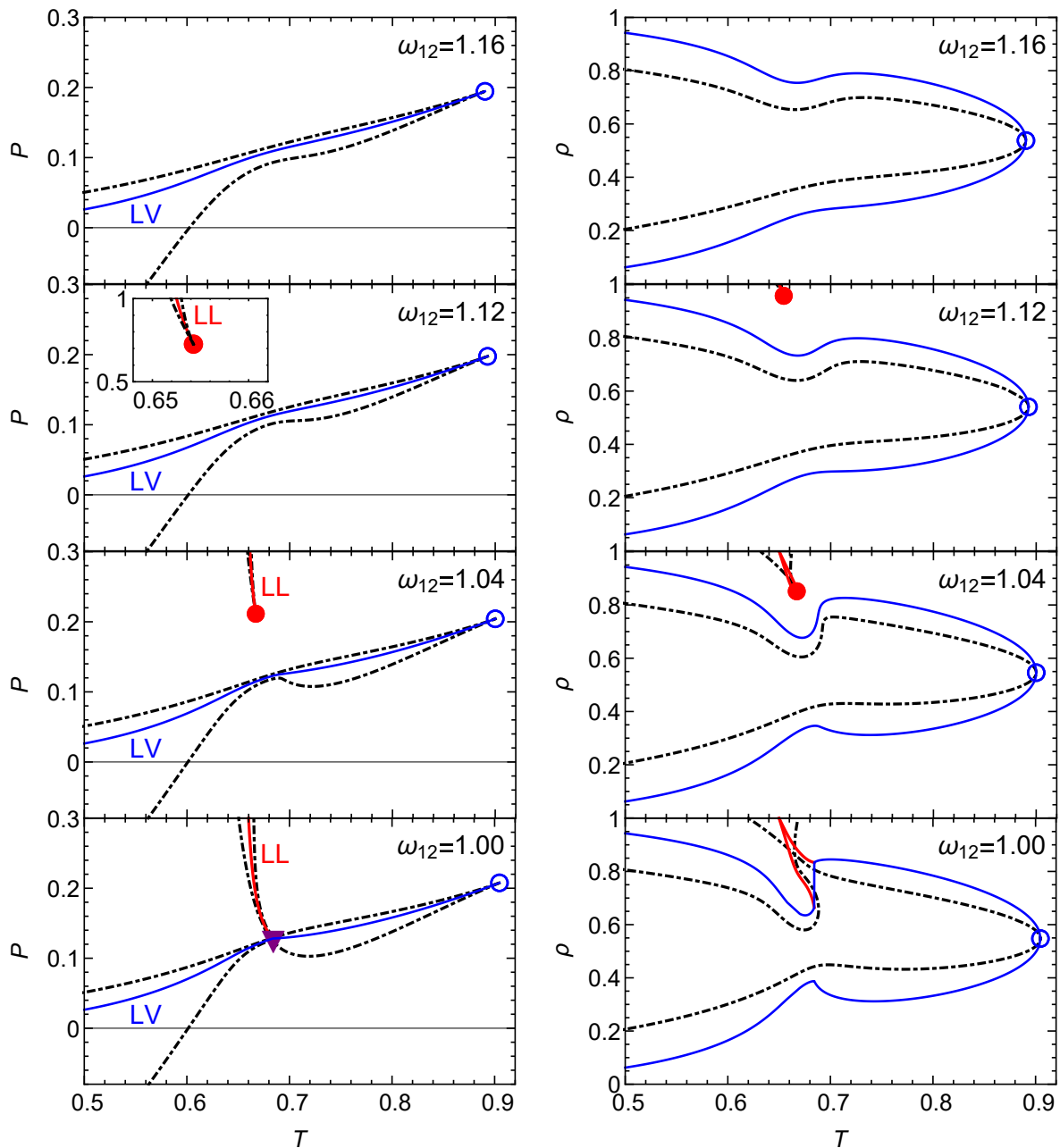}}
\caption{$T-P$ (left) and $T-\rho$ (right) phase diagrams of an interconverting fluid, for $\omega_1=1.6$, $\omega_2=2$, illustrating the four possible scenarii: singularity free ($\omega_{12}=1.16$), second critical point with monotonic spinodal ($\omega_{12}=1.12$), second critical point with non-monotonic spinodal ($\omega_{12}=1.04$), critical-point free ($\omega_{12}=1$). Curves shown are: liquid-vapor (LV, solid blue) and liquid-liquid (LL, solid red) equilibria; spinodals (dash-dotted black). Empty blue circle: LV critical point; filled red circle: LL critical point; purple triangle: triple point. The inset for $\omega_{12}=1.12$ shows the LL equilibrium occuring at higher pressure than displayed in the main graph.
\label{fig:scenarii}}
\end{figure*}

In order to understand the various scenarios that can be obtained for the interconverting fluid, a prerequisite is the knowledge of the underlying phase diagram for the non-interconverting, binary mixture. We use the classic compressible binary mixture on a lattice~\cite{trappeniers_gas-gas_1970,schouten_two-component_1974,furman_global_1977} (see Appendix for details). Consider a lattice whose sites can be either empty or occupied by only one particle of two species 1 and 2. The empty sites do not interact with the rest, whereas particles interact with their $z$ nearest-neighbors, with an interaction energy $-2 \omega_1 /z$, $-2 \omega_2 /z$, and $-2 \omega_{12}/z$, for $1-1$, $2-2$, and $1-2$ pairs, respectively. Figure~\ref{fig:binary3D} shows the $T-P-x$ phase diagram for a generic case, with $\omega_1 = 1.6$, $\omega_2 =2$, and $\omega_{12}=1.04$. At low temperature, liquid-liquid demixing occurs, with a liquid-liquid critical line (LLcl). This diagram belongs to an unusual case of Type II critical behavior in the classification of Konynenburg and Scott~\cite{van_konynenburg_critical_1980}, with in addition a re-entrant cusp in the $P-T$ projection of the critical line, making this case special~\cite{van_pelt_critical_1992}. $T-P$, $T-x$ and $T-\rho$ projections, where $\rho$ is the density, are displayed in Fig.~\ref{fig:binary2D}, together with those for the symmetric ($\omega_1 = \omega_2 =2$, $\omega_{12}=1.24$) and tricritical ($\omega_1 = 1.6$, $\omega_2 =2$, $\omega_{12}=1$) cases. Fig.~\ref{fig:binary3DTC} shows the $T-P-x$ phase diagram of the latter.

\section{One-component system with interconversion between states} 
To introduce LP, we allow the two species 1 and 2 to interconvert. In a binary system, the two chemical potentials for each pure species are independent, which means that adding a constant to one of them does not change the phase diagram nor thermodynamic properties. In contrast, with interconversion, the difference between the chemical potentials is restricted by the reaction equilibrium condition that depends on $T$ and $P$. We introduce this through the changes in energy, $e$, and in entropy $s$, when a particle changes state from 1 to 2 (see Appendix for details). This could for instance correspond to internal degrees of freedom which are frozen in state 1, but become accessible in state 2, which causes the number of internal configurations and hence the entropy in 2 to be higher than in 1. In the case of water, the model can be thought as a coarse grain model where a ``particle'' is a group of water molecules, who could be tetrahedrally arranged (state 1 with a lower energy and entropy) or more disordered (state 2 with a higher energy and entropy). This is reminiscent of the A and B states~\cite{anisimov_thermodynamics_2018} or $\rho$ and $\psi$ structures~\cite{tanaka_liquid-liquid_2020} in phenomenological models proposed for water. We note that as we do not need to specify the origin of the energy and entropy differences, our model is generic and can be applied to any type of polyamorphic fluid.

With interconversion, the system effectively becomes a single-component one, which follows specific $x(T,P)$ paths. Two examples are given in Fig.~\ref{fig:binary3D}. The first path shows liquid-vapor equilibrium in the interconverting fluid. The fraction $x$ decreases with increasing temperature, until a single liquid-vapor critical point is reached; it is located on the liquid-vapor critical line of the underlying binary phase diagram. The second path in Fig.~\ref{fig:binary3D} shows how the fraction changes with temperature in the interconverting fluid, along an isobar in the liquid region. For that particular choice ($P=0.18$), the path passes very close to the LLcl of the underlying binary phase diagram, but without crossing the liquid-liquid equilibrium surface. By tuning the model parameters, this crossing can be obtained for a range of pressures or avoided, either generating LP (e.g. a LLT with a liquid-liquid critical point (LLCP)) in the interconverting fluid, or not.

\begin{figure}[ttt]
\centerline{\includegraphics[width=\columnwidth]{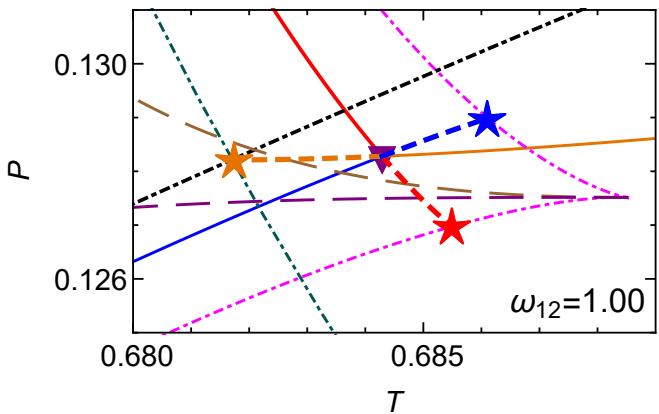}}
\caption{Close-up in the $T-P$ plane for the case $\omega_{12}=1$ displayed in Fig.~\ref{fig:scenarii}. The spinodals (dot-dashed curves) have been colored for clarity, as well as the three two-phase equilibrium lines (solid curves), whose metastable continuation is shown with short dashed curves. Their intersections with the spinodals define three ``Speedy points'' (stars). Two lines of anomalies (long-dashed) are also shown, see text and Fig.~\ref{fig:anoms} for details.
\label{fig:Speedy}}
\end{figure}

\begin{figure*}
\centerline{\includegraphics[width=0.9\textwidth]{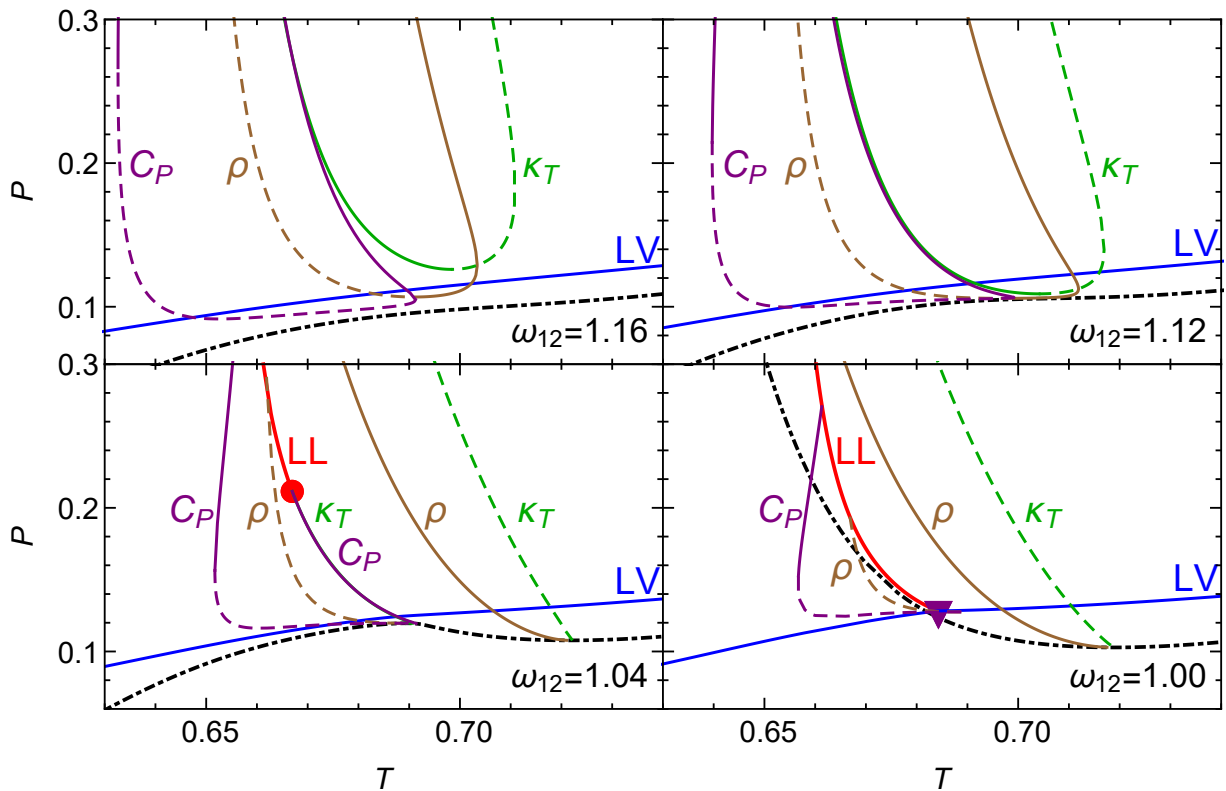}}
\caption{Lines of extrema in the $T-P$ plane for the four cases displayed in Fig.~\ref{fig:scenarii}. Solid and dashed curves show maxima and minima, respectively, of density ($\rho$, brown) and isothermal compressibility ($\kappa_T$, green) along isobars, and of isobaric heat capacity ($C_P$, purple) along isotherms. Also shown are the liquid-vapor equilibrium (LV, solid blue curve), the high density liquid spinodal (dot-dashed black curve), the LLCP (red disc), and the liquid-liquid-vapor triple point (purple triangle).
\label{fig:anoms}}
\end{figure*}

Figure~\ref{fig:scenarii} shows the various possible scenarios obtained by varying the non-ideal mixing parameter $\omega_{12}$, while keeping all other parameters constant ($\omega_1 = 1.6$, $\omega_2 =2$, $e=3$ and $s=4$). For $\omega_{12}>1.15$, there is only one liquid, with a liquid-vapor transition. The absence of a LLT is rigorously proven by studying the spinodal curves, whose temperature admits an analytic expression as a function of $x$ (see Appendix for details). The spinodals are physically acceptable only if they are located at densities below 1, the maximum possible value for the model when all sites are occupied. For $\omega_{12}>1.15$, only the liquid-vapor spinodals are acceptable. This case, with no LLT, corresponds to the singularity-free (SF) scenario~\cite{sastry_singularity-free_1996}. Figure~\ref{fig:scenarii} (top) illustrates this case for $\omega_{12}=1.16$: along the liquid-vapor equilibrium, the pressure is monotonic, whereas the density is not. This is the most prominent water-like anomaly, which is found for all cases shown in Fig.~\ref{fig:scenarii}. When $\omega_{12}$ is lowered below 1.15, a LLT appears, terminating in a LLCP. Three distinct cases are found. For $1.15>\omega_{12}>1.1075$, the LLCP is at relatively high pressure, and the liquid-vapor spinodal pressure is a monotonically increasing function of temperature; this corresponds to the ``second critical point scenario''~\cite{poole_phase_1992}. For $1.1075>\omega_{12}>1.018$, there is still a first-order LLT terminating in a LLCP, but the liquid-vapor spinodal pressure exhibits a maximum and a minimum as a function of temperature; this is a possibility that, to our knowledge, had not been proposed yet. Finally, for $\omega_{12}<1.018$, there is a LLT, but the LLCP disappears because it lies beyond the spinodal; this corresponds to the critical point free scenario~\cite{poole_effect_1994,angell_insights_2008}. In this last case, there is a triple point where two liquids and vapor coexist. Figure~\ref{fig:Speedy} shows how the metastable continuations of the LLT and the two liquid-vapor transitions each end when they touch the corresponding spinodal. The metastable equilibrium ceases because one of the phase becomes unstable. Our microscopic model thus confirms the findings of Ref.~\cite{chitnelawong_stability-limit_2019}, based on a phenomenological EoS, that a spinodal does not necessarily intersect a binodal at a critical point, but may terminate it at a so-called ``Speedy point''. This settles a 20-years old controversy~\cite{debenedetti_supercooled_2003,speedy_comment_2004,debenedetti_reply_2004} and demonstrates the viability of the critical-point free scenario.

\section{Lines of extrema of thermodynamic properties}
The vast majority of liquids shows a monotonic increase of molar volume, isothermal compressibility $\kappa_T$, and isobaric heat capacity $C_P$ when temperature increases along isobars. A liquid is anomalous when it exhibits extrema in these quantities, and water is considered to be the most anomalous liquid~\cite{gallo_water_2016}: along isobars, stable water shows maxima of $\rho$ and minima of $\kappa_T$, and maxima of $\kappa_T$ have been reported in metastable water~\cite{holten_compressibility_2017,kim_maxima_2017}. Our model captures such anomalies in the interconverting fluid, and Fig.~\ref{fig:anoms} shows their loci for each of the four cases displayed in Fig.~\ref{fig:scenarii}. The behaviour of $x$, $\rho$ and $\kappa_T$ in the four cases along the same isobar at $P=0.18$ is given in Fig.~\ref{fig:isobars}. The extrema lines follow a familiar pattern, obeying thermodynamic rules: when the lines of $\rho$ and $\kappa_T$ extrema along isobars intersect, the former reaches an extremum temperature~\cite{sastry_singularity-free_1996}, and when the lines of $\rho$ extrema along isobars and $C_P$ extrema along isotherms intersect, the former reaches an extremum pressure~\cite{poole_density_2005}. As shown in Fig.~\ref{fig:binary3D}, the anomalies observed in the interconverting fluid are related to the liquid-liquid equilibrium and critical line in the underlying binary fluid, whose distances to the $T-x$ path followed with interconversion varies with $P$ and $\omega_{12}$. The cases $\omega_{12}=1.16$ and $1.12$ are similar, with the loci of extrema avoiding the liquid-vapor spinodal which thus keeps a monotonic pressure~\cite{speedy_stability-limit_1982}. For $\omega_{12}=1.12$, Fig.~\ref{fig:anomszoom1} shows a close-up near the spinodal, and Fig.~\ref{fig:real} provides an enlarged view emphasizing the similarity with the lines of anomalies for real water.
In the cases with a LLCP, $\omega_{12}=1.12$ and $1.04$, lines of $\kappa_T$ maxima along isobars and of $C_P$ maxima along isotherms emanate from the LLCP. In the case $\omega_{12}=1.04$, the extrema in $\rho$, $\kappa_T$ and $C_P$ eventually intersect the liquid-vapor spinodal, causing it to go through extremum pressures~\cite{speedy_stability-limit_1982} (see Fig.~\ref{fig:anomszoom2} for a close-up near the maximum spinodal pressure). In the case $\omega_{12}=1$, the LLCP is pushed away in the unstable region. The line of maxima in $\kappa_T$ along isobars has disappeared; instead, $\kappa_T$ in the liquid metastable below the LLT diverges at the low temperature liquid-liquid spinodal. Figure~\ref{fig:Speedy} shows how the lines of minima in $\rho$ along isobars and in $C_P$ along isotherms extend into the metastable low-density liquid region, until simultaneously reaching its maximum temperature. There is now a continuous line of instability bounding the high-density liquid region at low pressure and at low temperature (Fig.~\ref{fig:anoms}), just as hypothesized by Speedy in 1982~\cite{speedy_stability-limit_1982}.

\section{Discussion}
Thermodynamics of LP has been mostly treated at the phenomenological level, based on an ad-hoc free energy specified with a modified van der Waals EoS~\cite{poole_effect_1994,chitnelawong_stability-limit_2019} or with mixing terms (two-state models)~\cite{tanaka_simple_2000,anisimov_thermodynamics_2018}. Attempts based on statistical mechanics to derive the free energy from a microscopic cell model exist~\cite{sastry_singularity-free_1996,franzese_intramolecular_2003,hruby_two-structure_2004,ciach_simple_2008,stokely_effect_2010,cerdeirina_waters_2019}. However, to generate the density maximum of water, they all introduce ``by hand'' a ``local'' density difference between the two states. Each cell changed occupancy~\cite{ciach_simple_2008} or volume~\cite{sastry_singularity-free_1996,stokely_effect_2010,cerdeirina_waters_2019} according to its actual state. Similarly, statistical mechanics and simulation approaches introduce two length scales~\cite{truskett_single-bond_1999,xu_relation_2005,gibson_metastable_2006,de_oliveira_ubiquitous_2009,vilaseca_isotropic_2011,pinheiro_critical_2017,urbic_modelling_2019,bartok_insight_2021} to describe water-like anomalies and LP, suggesting that the presence of two length scales is an ubiquitous ingredient for such phenomena.

These aspects have been a long-standing source of criticism against the application of two-state models to water, because the supposed large density contrast between the two states (e.g. 24\% ~\cite{ciach_simple_2008} or 38\% ~\cite{cerdeirina_waters_2019}) would be easily detected in x-ray or neutron scattering experiments~\cite{clark_small-angle_2010}. Our microscopic two-state model settles this debate. While generating the density anomaly and the possibility of LP, it does not require two different length scales: it is a fixed lattice in which each site can be occupied by only one particle without an explicit length or volume difference. We found that even in the perfectly symmetric case $\omega_1 = \omega_2$, for which the pure fluids under the same conditions would have exactly the same density, appropriate choices of $e$ and $s$ lead to water-like anomalies in the interconverting fluid. This underlines that non-ideality in the mixture is the primary ingredient for anomalous behaviour. This can be understood with Fig.~\ref{fig:binary2D} (top row): for instance, the path followed by $x$ as a function of temperature in the interconverting fluid can be tuned to cross the LLcl at a given pressure, thus generating the second critical point scenario and the associated anomalies. Even in the SF scenario without a LLT, the path followed by $x$ along liquid-vapor equilibrium can be tuned to pass close to the azeotrope \textbf{a}, before reaching the liquid-vapor critical line at higher temperature and higher density, which generates a non-monotonic density. In the more general case where $\omega_1 \neq \omega_2$ (Fig.~\ref{fig:binary2D}, middle row), the hypothetical pure fluids 1 and 2 have different densities under the same conditions of temperature and pressure; this further contributes to the density anomaly. This connects to purely two-state models such as in Ref.~\cite{anisimov_thermodynamics_2018} which specify differences between the hypothetical pure fluids 1 or 2, but do not imply a local density contrast in the interconverting fluid, i.e. bimodality of the distribution of local particle volumes. This resolves the controversy around the interpretation of two-state models: for instance, the ``locally favored structure'' of Ref.~\cite{tanaka_simple_2000} which has ``more specific volume than the normal-liquid structure'' should be understood as the average structure of the hypothetical pure component, rather than a local low-density structure in the interconverting fluid. 

In the case of water, it has been argued that the SF scenario could be obtained only in ``the artificial limit in which a molecule’s hydrogen-bonding connectivity is completely uncorrelated''~\cite{kim_maxima_2017}. This is indeed the case for a cell model studied by Stokely~\textit{et al.}~\cite{stokely_effect_2010} (where the SF scenario is obtained only for zero cooperativity between molecules forming hydrogen bonds), or in the phenomenological model of Ref.~\cite{anisimov_thermodynamics_2018} (where the SF scenario requires ideal mixing between the two states, $\omega=\omega_1+\omega_2-2\omega_{12}=0$). Here, we find that, even with significant nonideality in the interactions between the interconverting species, there could be a case with no LLT, corresponding to the SF scenario. Indeed, the LLT exists for large enough $\omega$, but as $\omega$ is lowered (i.e. $\omega_{12}$ increased), the LLCP, while staying at finite temperature, moves to higher density, until it reaches inaccessible states with density above 1 (Fig.~\ref{fig:scenarii}). More generally, depending on the details of the system, it is possible, with physically acceptable parameters and the presence of thermodynamic anomalies, that the LLT occurs in inaccessible regions of the phase diagram (e.g. in the \textit{non-thermodynamic habitat}~\cite{kiselev_parametric_2002} at which the metastable liquid has not enough time to equilibrate before crystallization occurs); such a case would be equivalent to the SF scenario. This might also solve the controversy about the existence of a LLT for the TIP4P/2005 water potential. Versions with reduced partial charges clearly show a LLT with LLCP~\cite{horstmann_relations_2021}. Simulations with the original TIP4P/2005 show critical-like behavior (at temperature above the putative LLCP) consistent with 3D Ising universality class~\cite{debenedetti_second_2020}, while advanced sampling techniques below the predicted LLCP temperature find an energy landscape with only one liquid phase~\cite{jedrecy_free_2021}. This could correspond to the interconverting fluid approaching closely the LLcl without crossing it. Importantly, we show that, for a given fluid, neither the shape of the line of density maxima, nor that of the liquid spinodal limit, nor the existence of $\kappa_T$ or $C_P$ maxima, is sufficient to identify which scenario is valid: a turning point in the line of density maxima, a monotonic liquid spinodal, or a line of $\kappa_T$ maxima along isobars, are found in cases with a LLT, as well as in cases without it.

The simplest case we considered, with fixed energy and entropy of interconversion, $e$ and $s$, already captures all scenarios proposed for water. Other fluids may exhibit different behaviors, such as a LLT with a positive $\mathrm{d}P/\mathrm{d}T$ slope as found for sulfur~\cite{henry_liquid-liquid_2020}, or terminated by two (lower and upper) LLCPs; this could be addressed using appropriate functions for $e$ and $s$. The interconverting lattice fluid thus provides a versatile tool to unify all types of anomalous fluids, with or without liquid polyamorphism.

\section*{Acknowledgments}
M.A.A. acknowledges the financial support of the National Science Foundation, award number 1856479, and F.C. that of Agence Nationale de la Recherche, grant number ANR-19-CE30-0035. We dedicate this work to the memory of late C. Austen Angell, our friend and mentor.

\appendix
\renewcommand{\thefigure}{A\arabic{figure}}

\setcounter{figure}{0}
\section*{Appendix\label{sec:app}}
\subsection{Compressible binary mixture in the canonical ensemble} 

In the 1970's, Trappeniers~\textit{et al.} introduced a variant of the lattice-gas model to describe a compressible binary mixture~\cite{trappeniers_gas-gas_1970,schouten_two-component_1974}. This two-component lattice-gas model was developped to account for the surprising finding of demixing in the gas phase or gas-gas equilibrium, observed in neon-krypton mixtures~\cite{trappeniers_gas-gas_1968}.

The system consists in a mixture of $N_1$ particles of type 1 and $N_2$ particles of type 2 placed among the $N$ sites of a rigid lattice with volume $V$, in contact with a thermostat at temperature $T$. Each site can be either empty or occupied by only one particle. The volume per lattice site is $v_0$, which we will set to $1$ in the following. The number density is thus $\rho=(N_1 + N_2 )/N$. The fraction of particles 1 in the mixture is $x=N_1 /(N_1 + N_2)$. A particle on a site interacts only with its $z$ nearest neighbors. There is no interaction with empty sites, and the interaction energy is $-\epsilon_1$, $-\epsilon_2$, and $-\epsilon_{12}$ for interactions between two particles of type 1, two particles of type 2, or between particles 1 and 2, respectively.

\begin{figure*}[ttt]
\centerline{\includegraphics[width=\textwidth]{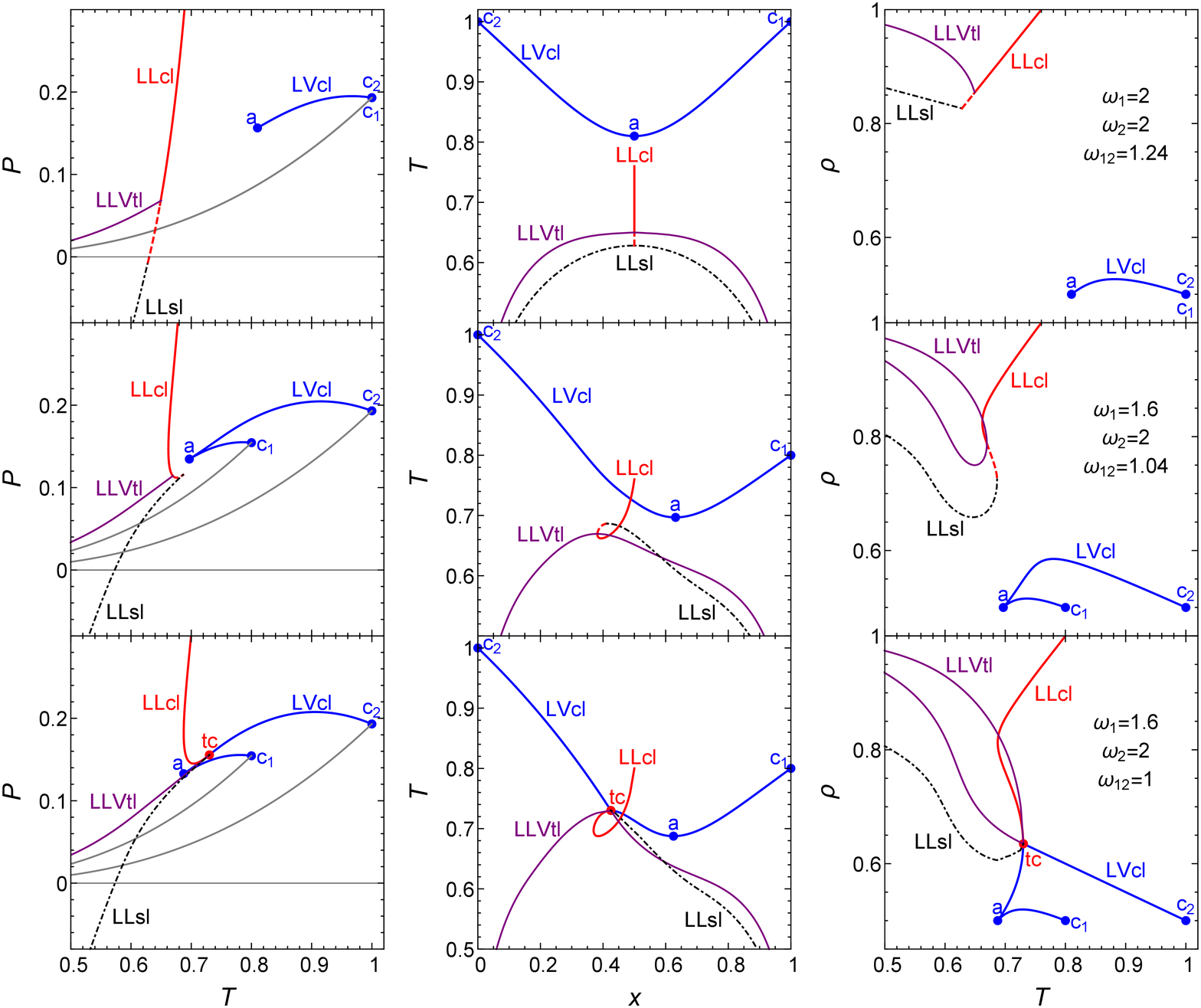}}
\caption{Projections of the phase diagram of a binary mixture in the $T-P$ (left), $x-T$ (middle), and $T-\rho$ (right) planes. Each line corresponds to a different set of interaction parameters, given in the right panels. The curves shown are: liquid-vapor equilibrium lines for the pure components (gray); critical lines relative to the liquid-vapor (LVcl, blue) and liquid-liquid (LLcl, red) transitions; triple lines (LLVtl, purple) where two liquids and one vapor coexist; spinodal lines for the liquid-liquid equilibrium (LLsl, dot-dashed black). The metastable part of the LLcl is shown by a dashed red curve. Salient points are: liquid-vapor critical points for the pure components (\textbf{c}$_1$ and \textbf{c}$_2$), azeotrope (\textbf{a}), and tricritical point (\textbf{tc}).
\label{fig:binary2D}}
\end{figure*}

\begin{figure*}
\includegraphics[width=0.85\textwidth]{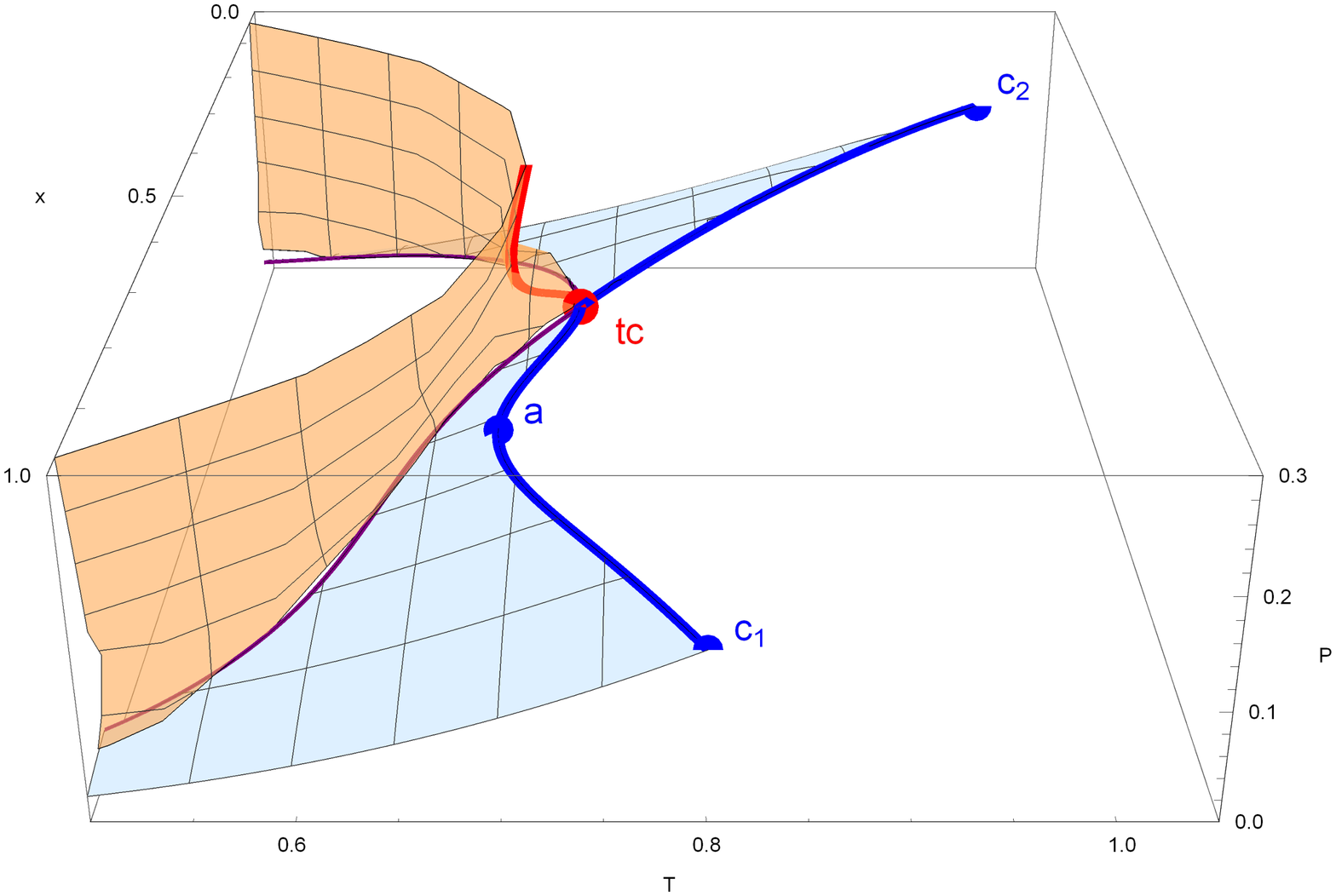}
\caption{$x-T-P$ phase diagram of a binary mixture with $\omega_1 = 1.6$, $\omega_2=2$, and $\omega_{12}=1$. The light blue surface shows the liquid in equilibrium with its vapor, terminating at the LVcl (blue) which reaches its minimum at the critical azeotrope \textbf{a}. The liquid-vapor critical points of the pure components are designated by \textbf{c}$_1$ and \textbf{c}$_2$. The orange surface shows the liquid-liquid equilibrium, terminating at the LLcl (red). The two surfaces intersect along a LLVtl (purple) where two liquids and one vapor coexist. The metastable parts of the liquid-liquid equilibrium surface and critical line are omitted for clarity. The liquid-liquid critical line ends on the liquid-vapor critical line at a tricritical point \textbf{tc}.
\label{fig:binary3DTC}}
\end{figure*}

In the following, we use the mean-field Bragg-Williams approach. Mean-field approximation is reasonably accurate, except for the immediate neighborhood of a critical point. Indeed, it is known that the physical properties, including the shape of phase boundaries in temperature-density coordinates, in the critical region of fluids, cannot be accurately described within the mean-field approximation. In particular, the exact values of the critical exponents differ from the mean-field values. The critical exponents are universal, being insensitive to the details of molecular interactions, since they arise from the long-range (mesoscopic) fluctuations of density and concentration. However, our work emphasizes the possibility of existence or absence of phase transitions. The shape of phase transition lines in pressure-temperature coordinates and the topology of the critical lines (in any variables) is determined by the short-range (microscopic) molecular interactions. This is why all the essential features of phase behavior of fluids and fluid mixtures can be realistically represented by mean-field approximation that remains to be the major tool for predicting or rejecting the existence of different phases.

The Helmholtz free energy per site can be written as
\begin{equation}
\begin{split}
f(T,\rho ,x) = &\frac{\Delta F}{N}\\
= & - \frac{z \rho^2}{2} \left[ \epsilon_1 x^2 + \epsilon_2 (1-x)^2 + 2 \epsilon_{12} \, x (1-x) \right] \\
& + \kB T \left\{ \rho  \left[ x \ln x + (1-x) \ln (1-x)\right]\right. \\
&+ \left. \rho \ln \rho + (1-\rho) \ln ( 1-\rho ) \right\} \, , \label{eq:f}
\end{split}
\end{equation}
where the first term is energetic and the second entropic. Here, $\Delta F$ is the free energy minus a temperature dependent function originating from the ideal-gas heat capacity; this function does not affect the phase equilibrium but is needed to calculate
the heat capacity. In the following we will set $\kB=1$. Introducing $\omega_1=-z \epsilon_1 /2$, $\omega_2=-z \epsilon_2 /2$, and $\omega_{12}=-z \epsilon_{12} /2$, $f$ can be rewritten as
\begin{equation}
\begin{split}
f(T, \rho , x) = & \rho \left\{ \rho \left[ \omega_1 x + \omega_2 (1-x) \right. \right. \\
& + \left. (2 \omega_{12} - \omega_1 - \omega_2 ) \, x (1-x) \right] \\
& + T \left. \left[ x \ln x + (1-x) \ln (1-x) \right] \right\} \\
& + T \left[ \rho \ln \rho + (1-\rho) \ln (1-\rho) \right] \, , \label{eq:f_bin_rhox}
\end{split}
\end{equation}

Following Schouten~\textit{et al.}~\cite{schouten_two-component_1974}, we introduce $\omega = \omega_1 + \omega_2 - 2 \omega_{12}$. It is easier to work with the variables $x_1=\rho x$ and $x_2 = \rho (1-x)$, which are the number densities of particles 1 and 2 in the system, respectively. The total number density $\rho$ and fraction $x$ are then given by $\rho=x_1 + x_2$ and $x = x_1/ (x_1 + x_2 )$. With the new variables, we obtain
\begin{equation}
\begin{split}
f(T,x_1 ,x_2)= & - (x_1 + x_2) ( \omega_1 x_1 + \omega_2 x_2 ) + \omega x_1 x_2 \\
& + T \left[ x_1 \ln x_1 + x_2 \ln x_2 \right. \\
& + \left. (1 - x_1 - x_2) \ln (1 - x_1 - x_2) \right]. \label{eq:f_bin_x1x2}
\end{split}
\end{equation}

The chemical potentials are then given by
\begin{align}
\begin{split}
\mu_1 = \left( \frac{\partial f}{\partial x_1}\right)_{T,x_2} = &- 2 \omega_1 x_1 - (\omega_1 + \omega_2 -\omega) x_2\\
&+ T \ln \frac{x_1}{1 - x_1 - x_2 }
\end{split}\\
\begin{split}
\mu_2 = \left( \frac{\partial f}{\partial x_2}\right)_{T,x_1} = &- (\omega_1 + \omega_2 -\omega) x_1 - 2 \omega_2 x_2\\
&+ T \ln \frac{x_2}{1 - x_1 - x_2 } \, ,
\end{split}
\end{align}
and the pressure by
\begin{equation}
\begin{split}
P = & - \left( \frac{\partial (Nf)}{\partial (N v_0)}\right)_{T,N x_1,N x_2} = -\frac{1}{v_0} \left( f - \mu_1 x_1 - \mu_2 x_2 \right) \\
= & -\frac{1}{v_0} \left[ \omega_1 {x_1}^2 + \omega_2 {x_2}^2 \right. \\
& + \left. (\omega_1 + \omega_2 - \omega ) \, x_1 x_2 + T \ln (1- x_1 -x_2 )\right]\, ,
\label{eq:P}
\end{split}
\end{equation}
where $v_0$ is the volume per lattice site, which we will set to $1$ in the following.

The system may exhibit several phases in equilibrium: liquid-vapor, liquid-liquid, or liquid-liquid-vapor coexistence. The set of equations for equilibrium between two phases A and B is
\begin{equation}
\mu_{1,\mathrm{A}}=\mu_{1,\mathrm{B}} , \, \mu_{2,\mathrm{A}}=\mu_{2,\mathrm{B}} , \, P_\mathrm{A}=P_\mathrm{B} \, .
\end{equation}
When three phases coexist, chemical potentials and pressure must be equal in all three phases, A, B, and C.

Introducing the notation $f_y = (\partial f / \partial y)$, we write the equation for the spinodal curve:
\begin{equation}
f_{x_1 x_1}f_{x_2 x_2} - f_{x_1 x_2}^2 =0 \, .
\end{equation}
Critical points fulfill an additional equation:
\begin{align}
\begin{split}
f_{x_1 x_1 x_1} \left( \frac{f_{x_1 x_2}}{f_{x_1 x_1}}\right)^3 - 3 f_{x_1 x_1 x_2} \left( \frac{f_{x_1 x_2}}{f_{x_1 x_1}}\right)^2 &\\
+ 3 f_{x_1 x_2 x_2} \left( \frac{f_{x_1 x_2}}{f_{x_1 x_1}}\right) - f_{x_2 x_2 x_2} &= 0 \, .
\end{split}
\end{align}
The liquid-liquid critical line tends to a maximum temperature $\omega/2$ when the density $\rho$ tends to $1$, the fraction $x$ to $1/2$, and the pressure to $+\infty$.

The above equations enable finding the phase diagram for the binary mixture. Different phase diagrams are possible depending on the values of $\omega_1$, $\omega_2$ and $\omega$. The master phase diagram in this parameter space is given by Furman~\textit{et al.}~\cite{furman_global_1977}. We discuss two particular cases below.

\subsection{Symmetric case} The symmetric case corresponds to $\omega_1 = \omega_2$. The free energy $f$ and the phase diagram become symmetric when exchanging $x$ and $1-x$, or $x_1$ and $x_2$. In particular, two liquid phases $A$ and $B$ in equilibrium will have fractions of particles 1 adding up to 1, i.e. $x_\mathrm{A}=x$ and $x_\mathrm{B}=1-x$, and the same density $\rho$. Consequently, the equation for the binodal reduces to the equation:
\begin{equation}
\begin{split}
\mu_{1,\mathrm{A}}=\mu_{1,\mathrm{B}} &= - 2 \omega_1 (x_1 + x_2) + \omega x_2 + T \ln \frac{x_1}{1 - x_1 - x_2 }\\
&= - 2 \omega_1 (x_1 + x_2) + \omega x_1 + T \ln \frac{x_2}{1 - x_1 - x_2 } \, ,
\end{split}
\end{equation}
because the solutions automatically fulfill the conditions $\mu_{2,\mathrm{A}}=\mu_{2,\mathrm{B}}$ and $P_\mathrm{A}=P_\mathrm{B}$. The binodal equation simplifies to
\begin{equation}
 \omega (x_1 - x_2) = T \ln \frac{x_1}{x_2} \, .
\end{equation}
With variables $\rho$ and $x$, this gives an analytic equation for the liquid-liquid binodal surface:
\begin{equation}
\rho = \frac{T}{\omega (1-2x)} \ln \frac{1-x}{x} \, .
\end{equation}
The pressure on the liquid-liquid binodal surface follows from Eq.~\ref{eq:P} as
\begin{equation}
P=[ \omega x (1-x) - \omega_1 ]\, \rho^2 - T \ln (1-\rho ) \, .
\end{equation}
Along the liquid-liquid critical line, $x=1/2$, $\rho=2T/\omega$, and the pressure varies as\begin{equation}
P_\mathrm{LLcl} = \left(\frac{T}{\omega}\right)^2 (\omega- 4\omega_1) - T \ln \left( 1 - \frac{2T}{\omega} \right) \, .
\end{equation}

At low (negative) pressure, the liquid-liquid binodal intersects the spinodal and the fluid becomes unstable. This happens where the liquid-liquid binodal along an isotherm reaches a minimum pressure. Because the pressure along the binodal is an analytic function of $T$ and $x$, finding its minimum along an isotherm gives a closed formula which allows expressing $T$, $P$, and $\rho$ as simple functions of $x$ along the curve defined by the intersection of the spinodal and the binodal surfaces. When $x=1/2$, this curve reaches a maximum temperature $(\omega/2)(8\omega_1 -\omega)/(8\omega_1 + \omega)$. At low (negative) pressure, the liquid-liquid binodal intersects the spinodal and the fluid becomes unstable. This happens where the liquid-liquid binodal along an isotherm reaches a minimum pressure. Because the pressure along the binodal is an analytic function of $T$ and $x$, finding its minimum along an isotherm gives a closed formula which allows expressing $T$, $P$, and $\rho$ as simple functions of $x$ along the curve defined by the intersection of the spinodal and the binodal surfaces. When $x=1/2$, this curve reaches a maximum temperature $(\omega/2)(8\omega_1 -\omega)/(8\omega_1 + \omega)$. The $T-P$, $T-x$ and $T-\rho$ projections are displayed in Fig.~\ref{fig:binary2D} (top row). According to the classification of Konynenburg and Scott~\cite{van_konynenburg_critical_1980}, the diagram belongs to Type II critical behavior: the vapor-liquid critical line is continuous, while the liquid-liquid critical line ends at the liquid-liquid-vapor triple line. However, since both components are equivalent in the individual states ($\omega_1 = \omega_2$), their liquid-vapor critical points coincide. The critical line turns back at the critical azeotrope where the composition of the mixture is $x = 0.5$. The liquid-liquid coexistence is symmetric with the liquid-liquid critical composition $x_\mathrm{c} = 0.5$.

\subsection{Tricritical case} When $\omega_1 = 1.6$, $\omega_2 =2$, and $\omega_{12}=1$, $\omega=\omega_1$ and there is ``asymmetric tricritical point''~\cite{rowlinson_liquids_1982}, as shown in Figs.~\ref{fig:binary3DTC} and~\ref{fig:binary2D} (bottom row), with two liquids and one vapor phase merging into the critical state. This is a limiting case of Type V critical behavior. In Type V the liquid-vapor critical line is separated by the liquid-liquid coexistence at two liquid-vapor critical end points. However, if the critical end points could merge into a single point, three coexisting phases become identical. Tricritical behavior is found in a number of three and four-component fluids~\cite{rowlinson_liquids_1982}, however, it has not yet been reported for binary systems since the tricriticality requires a very specific, if not unique, combination of the interaction parameters. As demonstrated in Fig.~\ref{fig:binary2D}, the phase diagram radically changed to the tricritical case upon only a slight decrease of the interaction parameter $\omega_{12}$ from 1.04 to 1.

\subsection{Case with interconversion\label{sec:inter}}

\paragraph{Formalism for interconversion}

We now introduce interconversion of the components in the general case. The fluid becomes a single component system with two accessible states. We impose a constant energy excess $e$ and a constant entropy excess $s$ for the state associated with type 2 particles, compared to the state associated with type 1 particles ($e$ and $s>0$). The Helmholtz free energy per lattice site, formerly Eq.~\ref{eq:f_bin_rhox}, is replaced by
\begin{equation}
\begin{split}
f(T,\rho ,x) = & - \rho x (e -T s)\\
& - \rho^2 \left[ \omega_1 x + \omega_2 (1-x) - \omega  x (1-x) \right] \\
& + T \left\{ \rho \left[ x \ln x + (1-x) \ln (1-x)\right] \right. \\
& + \left. \rho \ln \rho + (1-\rho) \ln ( 1-\rho ) \right\}\, . \label{eq:f_rhox}
\end{split}
\end{equation}

Equivalently, Eq.~\ref{eq:f_bin_x1x2} is replaced by
\begin{equation}
\begin{split}
f(T,x_1 ,x_2)= &-x_1 (e -T s) \\
& - (x_1 + x_2) ( \omega_1 x_1 + \omega_2 x_2 ) + \omega x_1 x_2 \\
& + T \left[ x_1 \ln x_1 + x_2 \ln x_2 \right. \\
& + \left. (1 - x_1 - x_2) \ln (1 - x_1 - x_2) \right]. \label{eq:f_x1x2}
\end{split}
\end{equation}

The equilibrium fraction $x$ is obtained by minimizing $f$:
\begin{equation}
\begin{split}
\left( \frac{\partial f}{\partial x}\right)_{T,\rho} = & - \rho \left\{ e - Ts + \rho \left[ \omega_1 - \omega_2 - \omega (1-2x) \right] \right\} \\
& + T \rho \ln \frac{x}{1-x} = 0 \, .\label{eq:xe}
\end{split}
\end{equation}

Interestingly, Eq.~\ref{eq:xe} gives the equilibrium density as a simple function of $T$ and $x$:
\begin{equation}
\rho = \frac{e-Ts -T \ln \left[ x/(1-x)\right]}{\omega_2 - \omega_1 + \omega (1-2x)}\, .\label{eq:rhoe}
\end{equation}

A peculiar case is obtained when the numerator and denominator both vanish, leading to an undeterminate $\rho$. This causes the crossing of all curves projected in the $(T,x)$ plane at
\begin{align}
T_{\times}&=\frac{e}{s+\ln \left[x_{\times}/(1-x_{\times})\right]} \, ,\\
x_{\times}&=\frac{1}{2}+\frac{\omega_2-\omega_1}{2\omega} \, .
\end{align}

Differentiating Eq.~\ref{eq:xe}, one gets
\begin{equation}
\rho \left(\frac{\partial x}{\partial \rho}\right)_T = \frac{\omega_2-\omega_1+\omega(1-2x )}{2\omega - T/\left[ \rho x (1-x) \right]} \, .\label{eq:dxedrho}
\end{equation}
The sign of this quantity is not trivial. However, at high enough temperature, $T>\omega/2$, the denominator is always negative, so that the sign is negative (resp. positive) when $x$ is below (resp. above) $x_{\times}$, and the fraction $x$ decreases (resp. increases) with increasing density.

The chemical potential is
\begin{equation}
\begin{split}
\mu = &\left( \frac{\partial f[T,\rho,x(T,\rho)]}{\partial \rho}\right)_T \\
= & \left( \frac{\partial f[T,\rho,x]}{\partial \rho}\right)_{T,x} \! [T,\rho,x(T,\rho)] \\
= & - x (e-Ts) - 2 \rho \left[ \omega_1 x + \omega_2 (1-x) - \omega x (1-x) \right] \\
& + T \left[ x \ln x + (1-x ) \ln (1-x ) + \ln \frac{\rho}{1-\rho} \right] \, .
\label{eq:mu}
\end{split}
\end{equation}

The pressure follows from Eqs.~\ref{eq:f_rhox} and \ref{eq:mu}:
\begin{equation}
\begin{split}
P=\mu \rho - f = & - \rho^2 \left[ \omega_1 x + \omega_2 (1-x) - \omega  x (1-x) \right]\\
& - T \ln ( 1-\rho ) \, .
\label{eq:Pinterconv}
\end{split}
\end{equation}
Phase equilibrium is found by solving conditions of equal pressure and equal chemical potential.

The critical points also satisfy $(\partial \mu/\partial \rho)_T=0$ and $(\partial^2 \mu/\partial \rho^2)_T=0$, which gives
\begin{widetext}
\begin{equation}
\begin{split}
& -2 \left[ \omega_1 x + \omega_2 (1-x) - \omega x (1-x) \right] + \frac{T}{\rho (1-\rho)} \\
& +2 \left\{ -(e - Ts) - 2 \rho [\omega_1 - \omega_2 -\omega (1-2x)] + T \ln \frac{x}{1-x}\right\} \left(\frac{\partial x}{\partial \rho}\right)_T
+\left[ - \omega \rho^2 + T \frac{\rho}{x (1-x)} \right] \left[\left(\frac{\partial x}{\partial \rho}\right)_T\right]^2 = 0 \,
\label{eq:spino}
\end{split}
\end{equation}
\end{widetext}
and
\begin{widetext}
\begin{equation}
\frac{2\rho - 1}{\rho^2 (1-\rho)^2} - 6 \rho [\omega_1 - \omega_2 -\omega (1-2x)]\left(\frac{\partial x}{\partial \rho}\right)_T + 3 \left[ - 2 \omega \rho + T \frac{1}{x (1-x)} \right] \left[\left(\frac{\partial x}{\partial \rho}\right)_T\right]^2 + \frac{2x - 1}{x^2 (1-x)^2} \left[\left(\frac{\partial x}{\partial \rho}\right)_T\right]^3 =0 \, .
\end{equation}
\end{widetext}

The spinodal curves correspond to the loci of points where $(\partial \mu/\partial \rho)_{T}$ vanishes. Eliminating $\rho$ between Eqs.~\ref{eq:rhoe},~\ref{eq:dxedrho}, and~\ref{eq:spino}, one obtains a cubic equation in $T$ whose roots give $T$ along the spinodal as a function of $x$.

\paragraph{Connection with phenomenological two-state models}

Instead of the Helmholtz free energy $f$, phenomenological two-state models~\cite{anisimov_thermodynamics_2018} usually work with the Gibbs free energy per molecule $G$, which is equal to $\mu$ in Eq.~\ref{eq:mu}. They introduce $G_{12}=G_1 - G_2$, the difference between the Gibbs free energies per molecule of entities 1 and 2, to write
\begin{equation}
G = G_2 + x G_{12} + T \left[ x \ln x + (1-x) \ln (1-x)\right] + \widetilde{\omega} x (1-x) \, ,\label{eq:Gibbs}
\end{equation}
where $\widetilde{\omega}$ is the parameter of non-ideality of mixing. It is useful to translate the parameters of the present model in the language of phenomenological two-state models. We have the following correspondence:
\begin{align}
G_2 & = - 2 \rho \omega_2 + T \ln \frac{\rho}{1-\rho} \, ,\\
G_{12}& = - T \ln K = - (e -Ts) - 2 \rho (\omega_1 - \omega_2) \, ,\\
\widetilde{\omega} & = 2 \rho \omega \, ,
\end{align}
where $K$ is the equilibrium constant for interconversion~\cite{anisimov_thermodynamics_2018}. Note that, to find the equilibrium fraction of interconversion $x$, $G$ in Eq.~\ref{eq:Gibbs} must be minizimed at constant temperature and pressure. After some algebra involving Eq.~\ref{eq:Pinterconv}, this leads to Eq.~\ref{eq:rhoe}.

\paragraph{Thermodynamic properties and their extrema}

\begin{figure}[ttt]
\centerline{\includegraphics[width=\columnwidth]{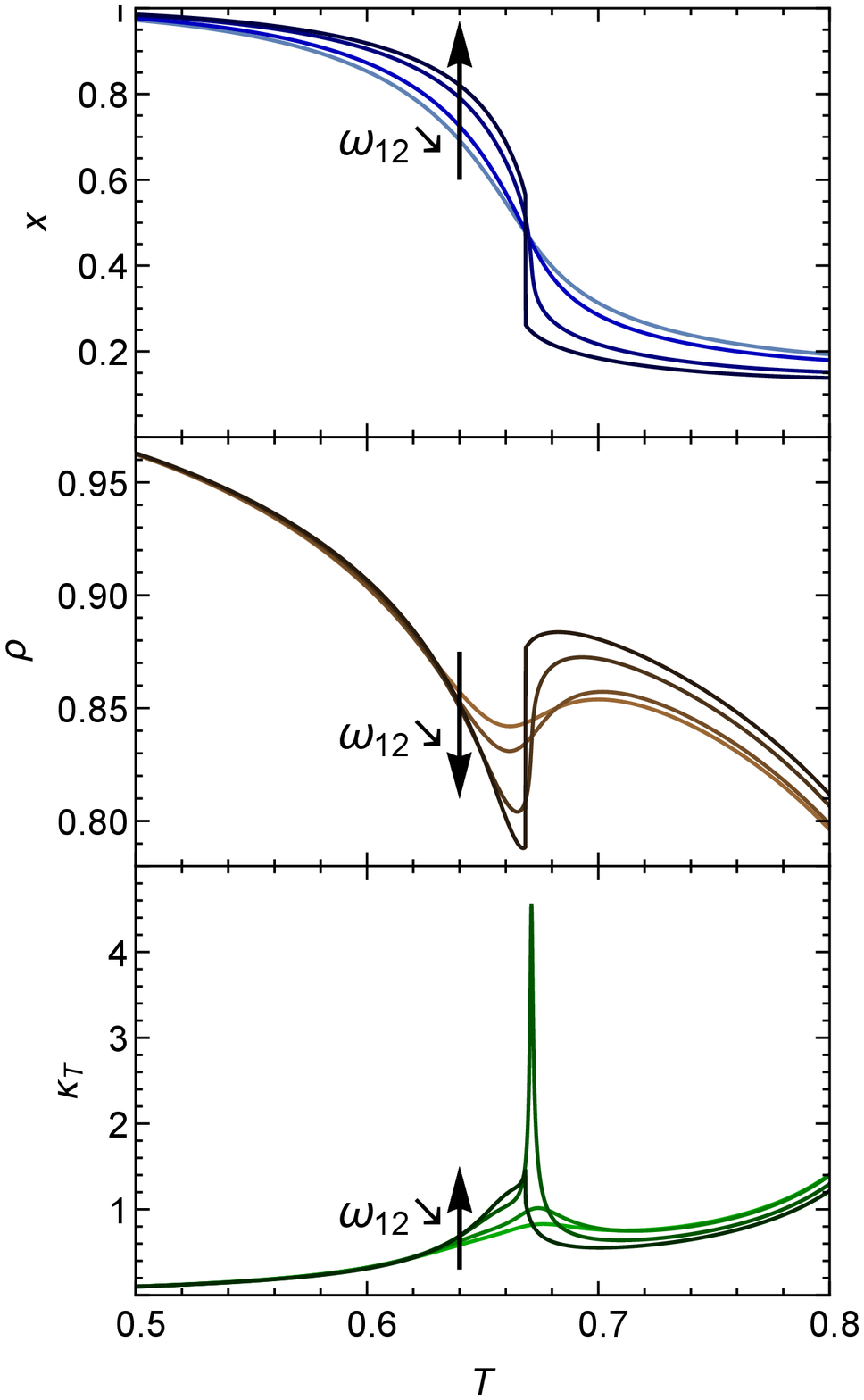}}
\caption{Fraction $x$ (top), density $\rho$ (middle), and isothermal compressibiity $\kappa_T$ (bottom) as a function of temperature along the $P=0.18$ isobar, for the four cases displayed in Fig.~\ref{fig:anoms}: $\omega_{12}=1.16$, $1.12$, $1.04$, and $1$. The arrows show directions of decreasing $\omega_{12}$, from lighter to darker curve. The darker curves ($\omega_{12}=1$) are discontinuous, whereas the others are continuous.
\label{fig:isobars}}
\end{figure}

\begin{figure}[hhh]
\centerline{\includegraphics[width=0.9\columnwidth]{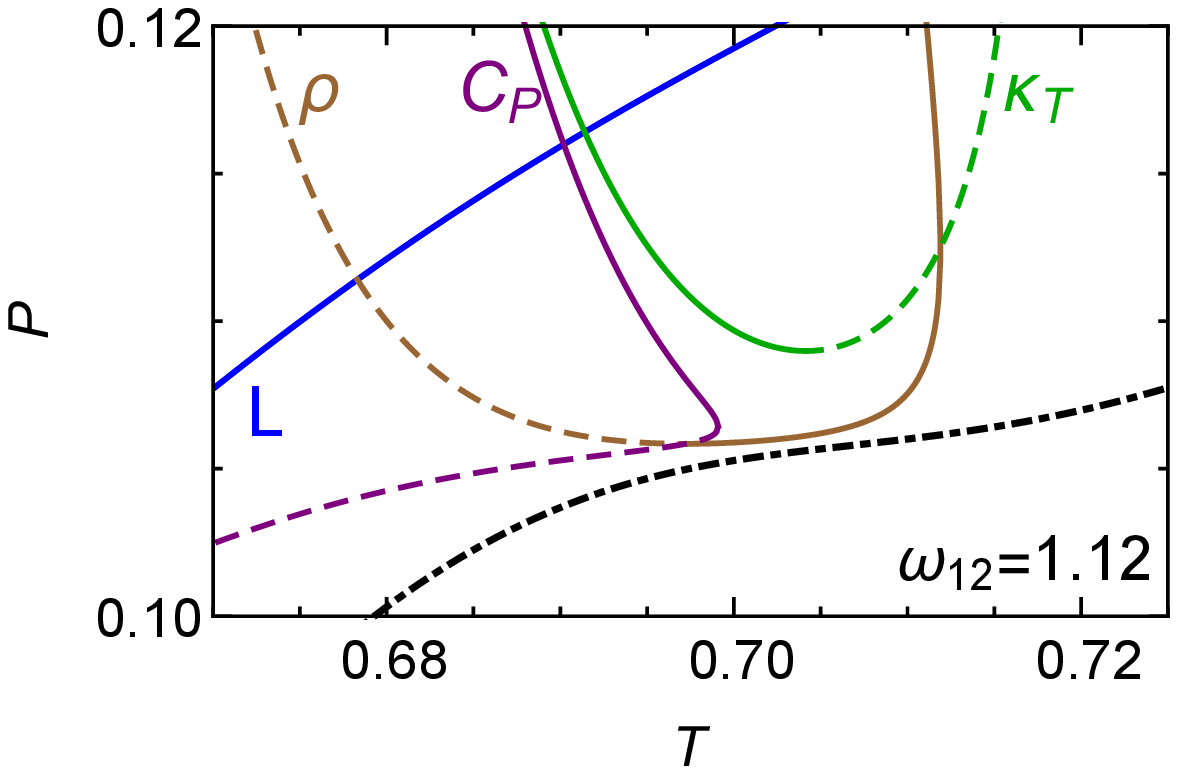}}
\caption{Close-up for the case $\omega_{12}=1.12$ displayed in Fig.~\ref{fig:anoms}. The lines of anomalies do not reach the spinodal that remains monotonic.
\label{fig:anomszoom1}}
\end{figure}

The density $\rho$ as a function of variables $T$ and $x$ is given by Eq.~\ref{eq:rhoe}. To find the density extrema along isobars, we write $\rho (T,x) =\rho [T, x(T,P)]$ and search for roots of
\begin{equation}
\begin{split}
\left(\frac{\partial \rho}{\partial T}\right)_{P} & = \left(\frac{\partial \rho}{\partial T}\right)_{x} + \left(\frac{\partial \rho}{\partial x}\right)_{T} \left(\frac{\partial x}{\partial T}\right)_{P} \\
& = 
\left(\frac{\partial \rho}{\partial T}\right)_{x} - \left(\frac{\partial \rho}{\partial x}\right)_{T} \left(\frac{\partial P}{\partial T}\right)_{x}
\left[ \left(\frac{\partial P}{\partial x}\right)_{T} \right]^{-1} \, ,
\label{eq:TMD}
\end{split}
\end{equation}
which can be written as a function of variables $T$ and $x$ only using Eqs.~\ref{eq:rhoe}, \ref{eq:mu}, and \ref{eq:Pinterconv}. To obtain the last equality in Eq.~\ref{eq:TMD}, use was made of the triple product relation:
\begin{equation}
\left(\frac{\partial x}{\partial T}\right)_{P} \left(\frac{\partial T}{\partial P}\right)_{x} \left(\frac{\partial P}{\partial x}\right)_{T} = -1 \, .
\label{eq:partials}
\end{equation}

To find extrema of isobaric heat capacity $C_P$ along isobars, we use the Maxwell relation:
\begin{equation}
\begin{split}
\left( \frac{\partial C_P }{\partial P} \right)_T & = -T \, \left( \frac{\partial^2 V}{\partial T^2} \right)_P \\
& =  \frac{T V}{\rho^2} \left\{ \rho \left(\frac{\partial^2 \rho}{\partial T^2}\right)_{P} - \left[ \left(\frac{\partial \rho}{\partial T}\right)_{P} \right]^2 \right\} \, .
\end{split}
\end{equation}
The term $(\partial \rho/\partial T)_P$ is given by Eq.~\ref{eq:TMD}, and the term $(\partial^2 \rho/\partial T^2)_P$ is obtained by differentiating Eq.~\ref{eq:TMD} and using Eq.~\ref{eq:partials}.

\begin{figure*}
\centerline{\includegraphics[height=11cm]{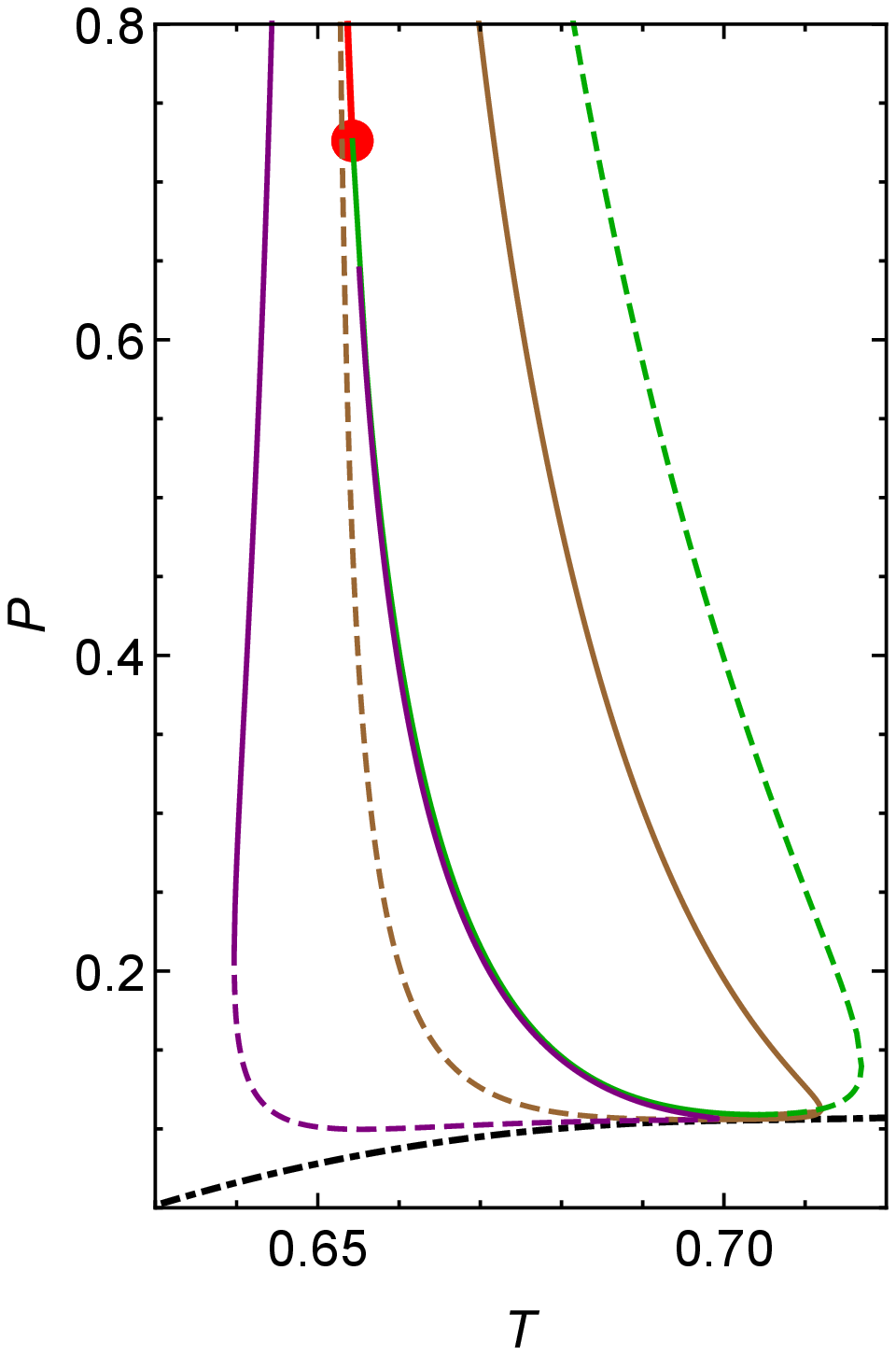}\hspace{10mm}\includegraphics[height=10.8cm]{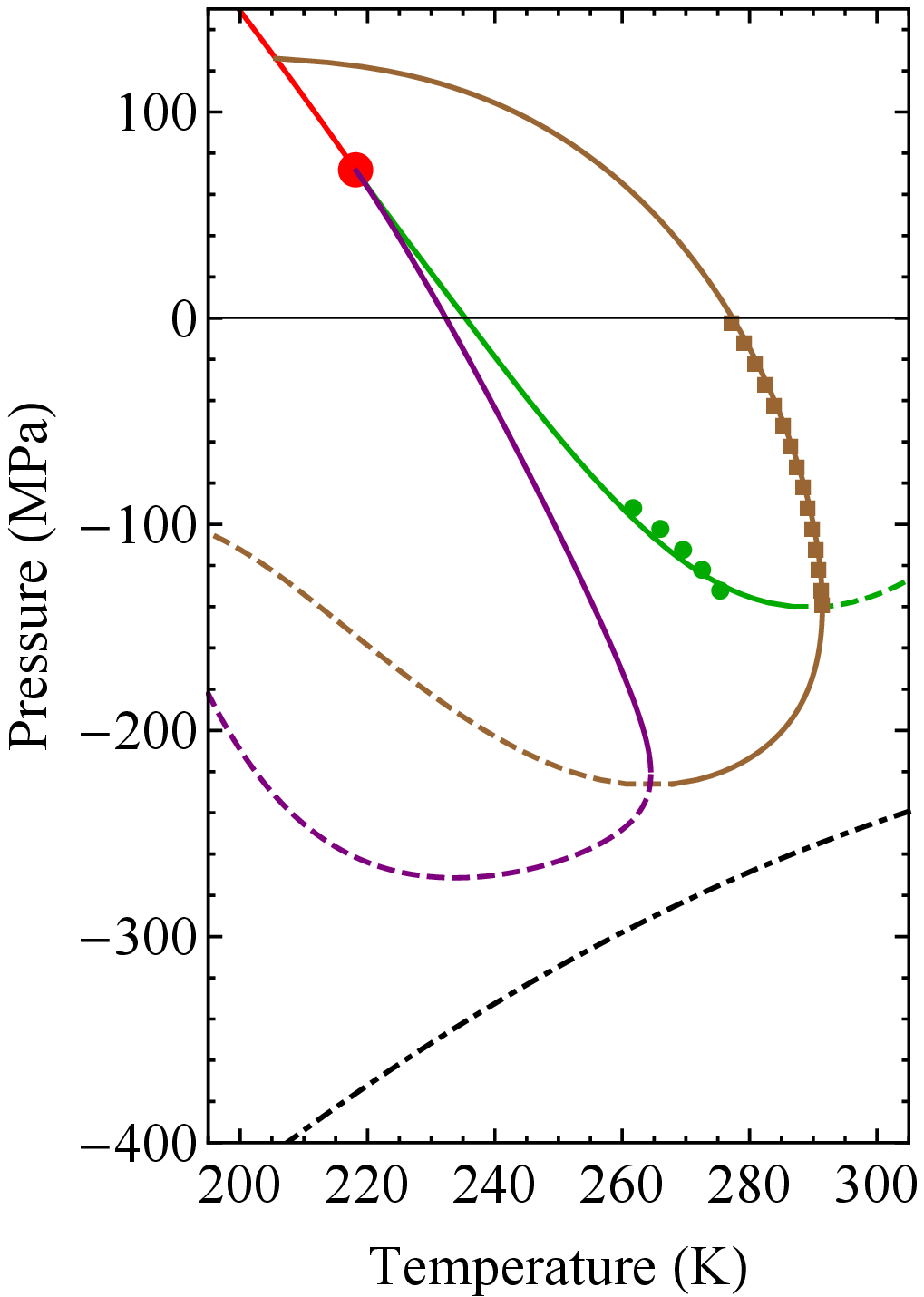}}
\caption{Left: enlarged view for the case $\omega_{12}=1.12$ displayed in Fig.~5 of the main text. Right: LLT (red curve), LLCP (red disc), spinodal (dot-dashed black curve) and lines of anomalies from the phenomenological model of Ref.~\cite{caupin_thermodynamics_2019}, which fits within their uncertainty 1098 experimental data point for real water with 21 adjustable parameters. This model reproduces the experimental lines of density maxima (brown squares) and of isothermal compressibility maxima (green discs) along isobars at negative pressure. The pattern of lines from the minimal model is qualitatively similar to that for real water, although its small number of parameters and simplicity do not allow for a quantitative match.
\label{fig:real}}
\end{figure*}

\begin{figure}[!]
\centerline{\includegraphics[width=0.95\columnwidth]{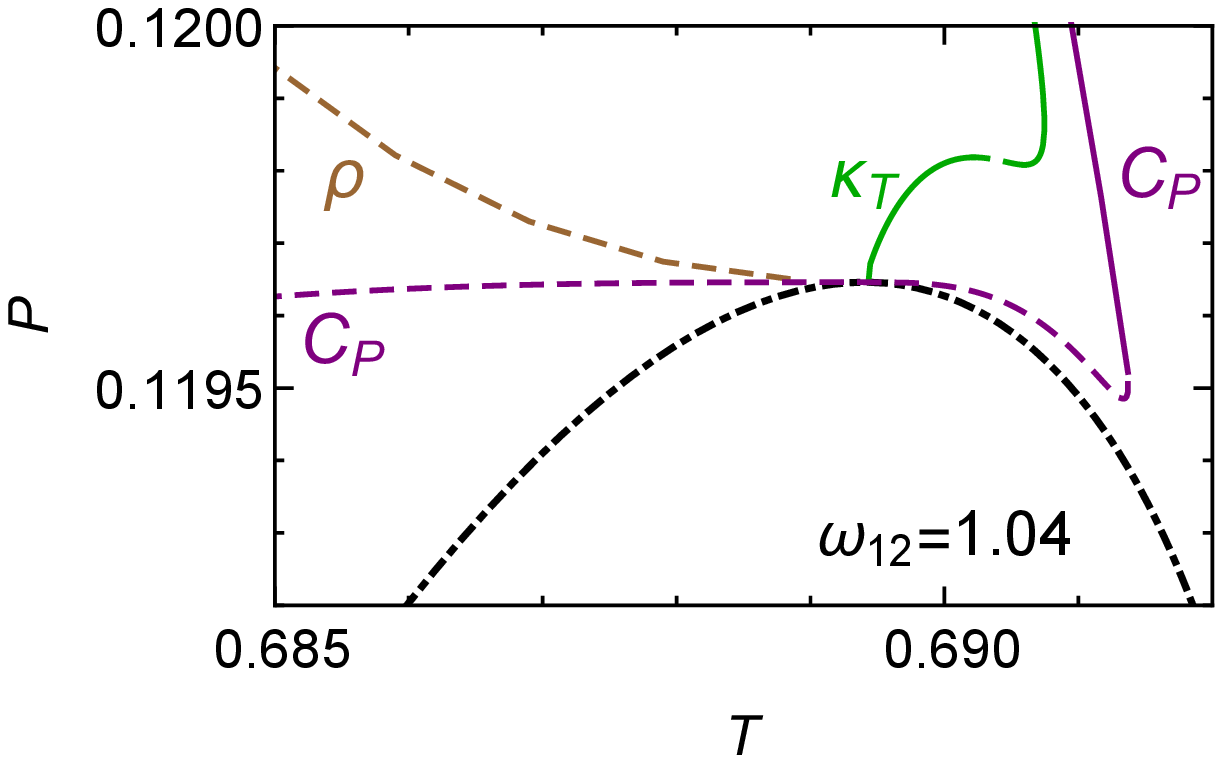}}
\caption{Close-up for the case $\omega_{12}=1.04$ displayed in Fig.~\ref{fig:anoms}. The lines of anomalies reach the spinodal, which exhibits a maximum (this figure) and a minimum (Fig.~\ref{fig:anoms}) in pressure.
\label{fig:anomszoom2}}
\end{figure}

\begin{figure}[!]
\centerline{\includegraphics[width=0.95\columnwidth]{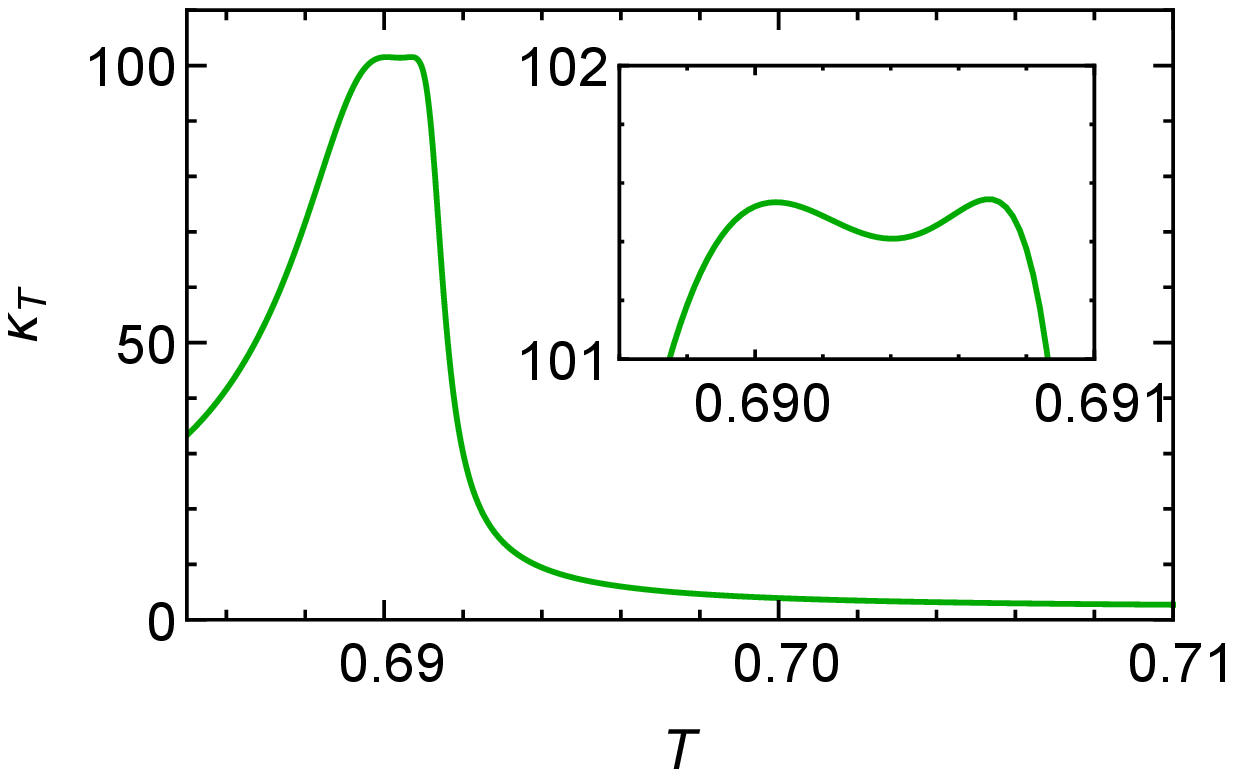}}
\caption{Isothermal compressibility $\kappa_T$ as a function of temperature $T$ along the isobar $P=0.119814$ for the case $\omega_{12}=1.04$ displayed in Fig.~\ref{fig:anoms}. The inset shows a close-up around the maximum to reveal the three extrema.
\label{fig:kTwiggle}}
\end{figure}
The isothermal compressibility $\kappa_T$ is obtained as
\begin{equation}
\kappa_T = \left[ \rho^2 \left(\frac{\partial \mu}{\partial \rho}\right)_T \right]^{-1} \, ,
\end{equation}
where the derivative is calculated using Eqs.~\ref{eq:dxedrho} and \ref{eq:mu}. To find extrema of $\kappa_T$ along isobars, the same reasoning as for Eq.~\ref{eq:TMD} is used with $1/\kappa_T$ instead of $\rho$.

Figure~\ref{fig:isobars} displays the behaviour of $x$, $\rho$ and $\kappa_T$ along the $P=0.18$ isobar for each of the four cases displayed in Figs.~\ref{fig:scenarii} and \ref{fig:anoms}.

The fraction $x$ always decreases with increasing temperature, as in Fig.~\ref{fig:binary3D}, faster for smaller $\omega_{12}$. For the curves corresponding to the lowest value $\omega_{12}=1$, the isobar crosses the liquid-liquid equilibrium surface of the binary phase diagram, which causes a discontinuity in $x$, $\rho$, and $\kappa_T$ (first-order LLT). Anomalies in thermodynamic functions are found along the $P=0.18$ isobar, with minima and maxima in $\rho$ and $\kappa_T$, except for $\omega_{12}=1$ for which the LLT occurs before a maximum in $\kappa_T$ is reached.

For $\omega_{12}=1.12$, Fig.~\ref{fig:anomszoom1} shows a close-up of the lines of anomalies near the spinodal, and Fig.~\ref{fig:real} (left) an enlarged view. We see that the model with $\omega_{12}=1.12$ gives a pattern of lines qualitatively similar to that for a model of real water~\cite{caupin_thermodynamics_2019} (Fig.~\ref{fig:real}, right). For $\omega_{12}=1.04$, Fig.~\ref{fig:anomszoom2} shows a close-up of the lines of anomalies near the spinodal, with a peculiar feature for $\kappa_T$: in a narrow pressure range (dashed portion of the green curve), $\kappa_T$ along isobars exhibits successively a  maximum, a minimum, and a maximum, as shown in Fig.~\ref{fig:kTwiggle}.


\bibliography{articles}

\end{document}